**A Calibrated Data-Driven Approach for Small Area Estimation using Big Data**

Siu-Ming Tam[1] and Shaila Sharmeen[2]

[1]National Institute of Applied Statistical Research, University of Wollongong, Northfields Avenue, Wollongong, NSW 2522, Australia and Methodology and Data Science Division, Australian Bureau of Statistics, ABS House, Benjamin Way, Belconnen 2617, Australia. Email: stattam@gmail.com.
[2]School of Information Technology, Deakin University, Burwood, Vic 3125, Australia. Email: shailacse2k@gmail.com.

Abstract

Where the response variable in a big data set is consistent with the variable of interest for small area estimation, the big data by itself can provide the estimates for small areas. These estimates are often subject to the coverage and measurement error bias inherited from the big data. However, if a probability survey of the same variable of interest is available, the survey data can be used as a training data set to develop an algorithm to impute for the data missed by the big data and adjust for measurement errors. In this paper, we outline a methodology for such imputations based on an kNN algorithm calibrated to an asymptotically design-unbiased estimate of the national total, and illustrate the use of a training data set to estimate the imputation bias and the "fixed - $k$ asymptotic" bootstrap to estimate the variance of the small area hybrid estimator. We illustrate the methodology of this paper using a public use data set and use it to compare the accuracy and precision of our hybrid estimator with the Fay-Harriot (FH) estimator. Finally, we also examine numerically the accuracy and precision of the FH estimator when the auxiliary variables used in the linking models are subject to under-coverage errors.

# 1    INRODUCTION

In official statistics, there is generally a significant, but often unmet, growing demand for small area statistics for decision making at the local level. By small area, we mean small geographical parcels of land where direct estimates from surveys generally fail to provide reliable estimates. Direct estimates fail because the sample size from probability surveys for the small areas is either too small or zero, i.e. there is no sample available for them, for reliable estimation. As a result, the "design-based approach" that is commonly adopted in official statistical surveys invariably fail to deliver good small area estimates (SAE). To overcome this problem, survey statisticians appeal to a "model-based approach" and use statistical models to "borrow strength" across areas (Fay and Herriot, 1979; Battese et al. 1988), across time (Pfeffermann and Tiller, 2006), or resort to synthetic estimation (Lehtonen and Veiganen, 2009). For a comprehensive account of the small area estimation methods, refer to Datta (2009), Ghosh and Rao (1994), Ghosh (2020), Jiang and Lahiri (2006a, 2006b), Pfeffermann (2002, 2013), Rao (2005, 2008) and Rao and Molina (2015).

With more and more data captured through the "internet of things (IoT)" - a system of interrelated computing devices, mechanical and digital machines, objects, animals or people that are provided with unique identifiers and the ability to transfer data over a network without requiring human-to-human or human-to-computer interaction (Daas et al., 2015; Tam and Clarke, 2015, Wiki, 2022) - what are the opportunities for SAE through borrowing strength from additional data sources? Whilst excellent attempts have been made in the literature to harness the information from the big data set, current research appears to be using the big data as a source for auxiliary variables in SAE unit level models (Battese et. al., 1988), or SAE area level models (Beaumout, 2020; Marchetti et al., 2015; Porter et al., 2014; Schmid et al., 2017; Rao, 2021). It is well known that many types of big data suffer from representational and measurement errors (Amaya et al., 2020), the use of big data as auxiliary variables in the Fay and Herriot (FH) model can lead to estimates which are worse than direct estimates (Ybarra and Lohr, 2008). Using a measurement error model, Ybarra and Lohr (2008) proposed a modified FH estimator by adjusting the weights of the convex combination of the direct estimate and model estimate of the small area, to address measurement errors in the auxiliary variables.

In this paper, we advocate a different approach to SAE by borrowing strength from big data. What is big data? According to the Big Data Privacy Report (Podesta *et al.*, 2014), the definition of big data depends on one's perspective - '…. there are many definitions of Big Data, which differ on whether you are a computer scientist, a financial analyst, or an entrepreneur pitching an idea to a venture capitalist…..'. The characteristics of big data have been popularly defined by five V's in the ICT literature, namely, Volume, Velocity, Variety, Veracity and Vulnerability. All of these do not actually define what big data is. Tam and van Halderen (2020) defined big data as the collection of the data from classical and IoT sources, and provided a schematic representation of the types of big data that may be useful to compile official statistics.

To illustrate ideas, we shall henceforth assume that there are no measurement errors in the big data, nor data linking errors between the big data and the probability survey data. The former issue was addressed in Tam et. al. (2020) and the latter in Kim and Tam (2021), and we shall not repeat their arguments here.

By considering the small area as comprising two strata, the big data stratum and a missing data stratum, the estimate for the small area is then the sum of the observed total of the variable of interest in the big data stratum and the estimated total in the missing data stratum. How do we estimate the total of the missing data stratum? If we have imputed values of the missing data, the total is then the sum of these imputed values. Imputing the missing values *en masse* is often referred to as mass imputation (Chipperfield et al., 2012). How do we impute? We can resort to classical methods (see, for example, Kim et. al., 2020) or machine learning algorithms. In this paper, we use a k-Nearest Neighbours (kNN) algorithm (Hastie and Tibshirani, p 463, 2008), but with the sum of the observed and imputed values calibrated to an approximately unbiased estimate - referred to as data integrator in Kim and Tam (2021) - of the population total at the national level. We call the resultant estimate as a hybrid estimate and the method hybrid estimation. The use of such calibration approach is not new - see for example, Beaumont (2005). Because the imputed values do not necessarily equal to the true but unobserved values, an imputation bias of the small area estimator is expected. Using the sample in the missing data stratum, we can estimate the imputation bias. Also inspired by the fixed - $k$ asymptotic method (Otsu and Rai, 2017) to bootstrap (Effron and Tibshirani, p. 124, 1986), we can estimate the variance of the hybrid estimator and thus its mean squared error to make inference on the small area total. Where the variable of interest in the big data is subject to measurement error, we can remove it using either a deterministic method if there is a priori information on the systematic nature of the error, or statistical modelling if the error is random. For the rest of the paper, we assume that the variable of interest is measured without error in both the big data and probability survey data sets.

The layout of the paper is as follows. The notation will be introduced and some well-known results are presented in Section 2. The methodology for SAE using big data and kNN subject to calibration – hereinafter referred to as a calibrated kNN algorithm (CkNN), imputation bias and confidence interval estimation will be described in Section 3. Section 4 gives an application of the methodology to a population data set from the Australian Bureau of Statistics and reports the comparison of the CkNN small area estimates with the FH estimates. It also examines the quality of the FH estimates when the auxiliary variables derived from big data are subject to coverage error. Finally, we conclude in Section 5.

## 2　　NOTATIONS AND AN ESTABLISHED RESULT

Suppose we have a finite population, $U = \{1,...,N\}$ comprising $N$ units with the following values, $\boldsymbol{x}_i$ and $y_i$, $\forall i \in U$, where $\boldsymbol{x}_i$ is a vector of auxiliary variables and is fully observed, and $y_i$ is the variable of interest. We assume that $U = B \cup C$, where $B$, of size $N_B$, comprises the labels of the big data set and $C$, of size $N_C$, is the complement of $B$. We assume further that $y_i$, $\forall i \in B$, are observed without error. Finally, we also assume that we have a probability sample, $A \subset U$, with known design weights of the sample, $d_i$, $\forall i \in A$. Thus we have the following data available to the analyst for SAE: (a) $(\boldsymbol{x}_i, y_i)$ for $i \in B$; (b) $(d_i, \boldsymbol{x}_i, y_i)$ for $i \in A$; (c) $\boldsymbol{x}_i$ $\forall i \in C$ and (d) information on where these units are located in the small area. Finally, let $\delta_i$ denote the big data inclusion indicator which is 1 if unit $i \in B$ and 0 otherwise. We assume that (e) $\delta_i$, $\forall i \in A$, is fully observed. Note that $\delta_i = 1$ is observed for $\forall i \in B$. In addition, note that the case when $A$ is subject to nonresponse and $\delta_i$ not fully observed was addressed in Tam et al (2020) and Kim and Tam (2021) respectively and will not be repeated here.

Suppose further that $U = U_1 \cup ... U_m \cup ... \cup U_M$, $B = B_1 \cup ... B_m \cup ... \cup B_M$ and $C = C_1 \cup ... C_m \cup ... \cup C_M$ and $A = A_1 \cup ... A_m \cup ... \cup A_M$, where $U_m = B_m \cup C_m$ and $m$ denotes the $m^{th}$ small area. For SAE, we are interested to estimate $T_m = \sum_{i \in U_m} y_i, m = 1,...,M$. As

$T_m = \sum_{i \in B_m} y_i + \sum_{i \in A_m \setminus B_m} y_i + \sum_{i \in C_m \setminus A_m} y_i = T_{B_m} + T_{A_m \setminus B_m} + T_{C_m \setminus A_m}$, and because $T_{B_m}$ and $T_{A_m \setminus B_m}$ are fully observed, the SAE problem boils down to estimating $T_{C_m \setminus A_m}$, using the information available from (a) to (e) above.

Denote the population total by $T = \sum_{m=1}^{M} T_m$. Kim and Tam (2021) showed that the data integrator,

perhaps better referred to as a hybrid estimator, $\hat{T}_P = \sum_{i \in U} \delta_i y_i + N_C \dfrac{\sum_{i \in A} d_i (1 - \delta_i) y_i}{\sum_{i \in A} d_i (1 - \delta_i)}$, is equivalent to a generalised regression estimators and hence is approximately designed unbiased (Särndal et. al., 1992, p 235). For simple random sampling of size $n$, $Var(\hat{T}_P) \approx (1 - W_B)\dfrac{N^2}{n}S_C^2$, where

$W_B = N_B / N$, $n / N \approx 0$, $S_C^2 = N_C^{-1}\sum_1^N (1 - \delta_i)(y_i - \bar{Y}_C)^2$ and $\bar{Y}_C = N_C^{-1}\sum_1^N (1 - \delta_i) y_i$. Furthermore,

if $S^2 = N^{-1}\sum_1^N (y_i - \bar{Y})^2$ and $\hat{T}_A = N\sum_{i \in A} y_i / n$, Kim and Tam (2021) showed that

$\dfrac{Var(\hat{T}_P)}{Var(\hat{T}_A)} = (1 - W_B)\dfrac{S_C^2}{S^2} < 1$, if $(1 - W_B)S_C^2 < S^2$. In other words, when $W_B$ is sufficiently large or

$S_C^2 \approx S^2$, $\hat{T}_P$ is a more efficient estimator than $\hat{T}_A$. As $\hat{T}_P$ is an efficient estimator of the population

total, we use it as the constraint to mass impute the missing values in $C \setminus A$.

## 3 METHODOLOGY FOR SAE WITH BIG DATA

Let $\hat{y}_i$ denote the imputed value of the missing data for unit $i \in C$. We want the imputed values to satisfy the following conditions:

(a) $\hat{y}_i = y_i$ if $i \in D$ where $D = A \cap C$; and

(b) $\hat{T}_C = \hat{T}_P - T_B$, where $\hat{T}_C = \sum_{i \in C} \hat{y}_i$ and $T_B = \sum_{i \in B} y_i$.

Provided that such an imputation methodology is developed, we can estimate $T_m$ by

$\hat{T}_m = T_{B_m} + \hat{T}_{C_m} = T_{B_m} + \sum_{i \in C_m} \hat{y}_i$. Here, $\hat{T}_m$ can be considered as a synthetic SAE. From (b),

$\hat{T} = \sum_{m=1}^M \hat{T}_m = \sum_{m=1}^M (T_{B_m} + \hat{T}_{C_m}) = \hat{T}_P$. Subjecting small area estimates to constraints is not a new concept, see, for example, Pfeffermann and Tiller (2006). In their paper, the constraint is used to provide protection against possible model failure. In this paper, the constraint is imposed to ensure that the sum of small area estimates equals to the efficient and approximately unbiased estimator of the national population total, $\hat{T}_P$, to ensure consistency between the sum of the SAEs and national total. This approach is also used in Beaumont (2005).

### 3.1 DEVELOPING THE CALIBRATED kNN (CkNN) ALGORITHM

Because $D = A \cap C$, is a probability sample of $C$, i.e. each unit in $C$ has a known and non-zero probability of inclusion in $D$, the MAR assumption and positivity assumptions to justify the use of kNN (Yang and Kim, 2019) are satisfied. Let $D = D_1 \cup ... D_m \cup ... \cup D_M$ and $|D_m| = n_{D_m}$ where $n_{D_m}$ is known. We introduce a second subscript $m$ to denote the small area in which the training data point is located, e.g. $y_{mi}$ denotes that the data point $i$ is located in small area $m$.

For the kNN algorithm, there are two hyper-parameters to be estimated, namely, the number of nearest neighbours, $k$, and number and nature of the features, $p = \dim(\boldsymbol{x}_i)$ to be determined using "feature engineering" as so described in data science or variables selection in statistics. Mathematically, we want to find the optimum $k$ and $p$ such that sum over $M$ of the absolute prediction error in each small area is a minimum. If $\hat{y}_{mi}$ (which depends on $\boldsymbol{x}_{mi}, k, p$) denotes the predicted value of $y_{mi}$, $L(\hat{y}_{mi}, y_{mi}) = \hat{y}_{mi} - y_{mi}$, denotes the "loss function" being defined here as the prediction error for the data point, the prediction error for the $m^{th}$ small area is $\sum_{i=1}^{n_m} (\hat{y}_{mi} - y_{mi})$, where $| C_m \setminus A_m | = n_{C_m \setminus A_m}$.

Following Wesley et al (2022) and Hastie et al (2008, p. 181), we use K-fold cross validation applied to the training data set $D$, to determine $k$ and $p$, which are to be solutions to minimise the following objective function:

$$\arg\min_{k,p} \left\{ \frac{1}{n_D} \sum_{j=1}^{K} \sum_{m=1}^{M} | \sum_{mi:\rho(mi)=j} L\left(y_{mi}, f^{-j}(\boldsymbol{x}_{mi}, k, p)\right) | \right\} \qquad (1)$$

where $n_D$ is the sample size of $D$; $\hat{y}_{mi} = f^{-j}(\boldsymbol{x}_{mi}, k, p)$ is the kNN predictor of $y_{mi}$, given the auxiliary vector, $\boldsymbol{x}_{mi}$, $k$ and $p$ based on the data from $D$ less the $j^{th}$ fold; and $\rho : \{1,...,n_D\} \mapsto \{1,...,K\}$ is the indexing function that indicates the fold to which observation indexed by $mi$ is allocated by a randomization process in such a way that the folds are as far as possible equal in size, so that the third summation sign in (1) denotes the summation over those $mi$'s in the $j^{th}$ fold. The objective function $\left\{ \frac{1}{n_D} \sum_{j=1}^{K} \sum_{m=1}^{M} | \sum_{mi:\rho(mi)=j} L\left(y_{mi}, f^{-j}(\boldsymbol{x}_{mi}, k, p)\right) | \right\}$ is referred to as the estimated test error rate (Hastie et al, p. 181, 2008) from the training data set. Note that we want the kNN algorithm trained to minimise $\sum_{m=1}^{M} | \sum_{i=1}^{n_{J_m}} (\hat{y}_{mi} - y_{mi}) |$ rather than $\sum_{m=1}^{M} \sum_{i=1}^{n_{J_m}} (\hat{y}_{mi} - y_{mi})$ for SAE, where $n_{J_m} = | J_m |$ and $J_m$ represents the $m^{th}$ small area in the fold. This is because we want to minimise the prediction error in all of the small areas, and do not want the error in one area to be offset by the errors of other areas. This will be an appropriate objective function as long as we are interested in getting the predicted counts in the small areas as close as possible to the actual count, and not accuracy in the individual predicted values, in which case, the objective function would be $\sum_{m=1}^{M} \sum_{i=1}^{n_{J_m}} | (\hat{y}_{mi} - y_{mi}) |$.

What distance metric should be used to find the nearest neighbours? Alfeilat et al. (2019) gave a comprehensive review and evaluation of the distance metrics that may be used in kNN algorithms and concluded that the Hassanat Distance (HasD) metric (Hassanat, 2014) performs the best when applied to a diversity of data sets. The HasD, which is used in the application in Section 4 below, between the points $\boldsymbol{x}_i = (x_{i1}, \ldots, x_{ip})^T$ and $\boldsymbol{x}_j = (x_{j1}, \ldots, x_{jp})^T$, is defined as follows:

$$\text{HasD}(\boldsymbol{x}_i, \boldsymbol{x}_j) = \frac{1}{p} \sum_{l=1}^{p} D(x_{il}, x_{jl}) \text{ where for } l = 1, \ldots, p, \text{ and } x_{il} \text{ or } x_{jl} \neq 0$$

$$D(x_{il}, x_{jl}) = 1 - \frac{1 + \min(x_{il}, x_{jl})}{1 + \max(x_{il}, x_{jl})}, \text{ if } \min(x_{il}, x_{jl}) \geq 0; \text{ or } 1 - \frac{1 + \min(x_{il}, x_{jl}) + |\min(x_{il}, x_{jl})|}{1 + \max(x_{il}, x_{jl}) + |\min(x_{il}, x_{jl})|} \text{ otherwise.}$$

Where $x_{il} = x_{jl} = 0$, we define $D(x_{il}, x_{jl}) = 0$. Furthermore, Hassanat (2014) showed that $D(x_{il}, x_{jl})$ is bounded between 0 (when $x_{il} = x_{jl} \forall l = 1, \ldots, p$) and 1 (when the distance $x_{il}$ and $x_{jl}$ is infinite for one $l = 1, \ldots, p$); symmetric (i.e. $D(x_{il}, x_{jl}) = D(x_{jl}, x_{il})$) and satisfies the triangular inequality (i.e. $D(x_{il}, x_{jl}) \leq D(x_{il}, x_{kl}) + D(x_{kl}, x_{jl})$).

Under kNN, the predictor (i.e. imputed value) is the arithmetic average of the k nearest neighbours, i.e. the weight attached to each of the nearest neighbours is the same $\frac{1}{k}$. In other words, if $\{y_{i(1)}, \ldots y_{i(j)}, \ldots, y_{i(k)}\}$ is the set of the $k$ nearest neighbours of $i \in C \setminus D$ as determined by HasD, the kNN predictor is $\sum_{j=1}^{k} \frac{y_{i(j)}}{k}$. Instead of using an arithmetic average of the $k$ nearest neighbours, we use a convex weighted average so as to ensure that the imputed values satisfy the calibration property (see (a) and (b) in the Lemma below), and want the weights, $w_j, j = 1, \ldots, k$, to be chosen as close to $\frac{1}{k}$ as possible. We use the Chi-square distance of Deville and Sarndal (1992) to determine closeness, i.e. $\sum_{j=1}^{k} k(w_j - \frac{1}{k})^2$.

**Lemma.** The solution to $\arg \min_{w_j} \sum_{j=1}^{k} \frac{1}{k}(kw_j - 1)^2$ subject the following constraints:

a.  $\sum_{i \in C} \hat{y}_i = \hat{T}_p - T_B = \hat{T}_C$, where $\hat{y}_i = y_i$ for $i \in D$, and $\hat{y}_i = \sum_{j=1}^{k} w_j y_{i(j)}$ for $i \in C \setminus D$; and

b.  $\sum_{j=1}^{k} w_j = 1$

is:

$$w_j = \frac{1}{k} + \frac{\left\{ \hat{T}^{(j)} - \frac{1}{k} \sum_{j=1}^{k} \hat{T}^{(j)} \right\}}{\left\{ \sum_{j=1}^{k} (\hat{T}^{(j)})^2 - \frac{1}{k} (\sum_{j=1}^{k} \hat{T}^{(j)})^2 \right\}} (\hat{T}_P - \hat{T}_{kNN}) \qquad (2)$$

where $\hat{T}^{(j)} = \sum_{i \in C \backslash D} y_{i(j)}, j = 1, ..., k$, $\hat{T}_{kNN} = T_B + T_D + \frac{1}{k} \sum_{j=1}^{k} \hat{T}^{(j)}$, and $T_D = \sum_{i \in D} y_i$. Thus

$\hat{T}_{CkNN} = T_B + T_D + \sum_{j=1}^{k} w_j \hat{T}^{(j)}$, where $w_j$ is given by (2).

**Proof**

The constraint in (a) can be rewritten as $\hat{T}_C - T_D = \sum_{i \in C \backslash D} \sum_{j=1}^{k} w_j y_{i(j)} = \sum_{j=1}^{k} w_j \hat{T}^{(j)}$. Using Lagrange

Multipliers, $2\lambda_1$ for (1) and $2\lambda_2$ for (b), differentiating $\sum_{j=1}^{k} k(w_j - \frac{1}{k})^2 = \sum_{j=1}^{k} \frac{1}{k}(kw_j - 1)^2$ and

simplifying, we have:

$$w_j = \frac{1}{k} + \frac{\lambda_1 \hat{T}^{(j)}}{k} + \frac{\lambda_2}{k}. \qquad (3)$$

Using $\hat{T}_C - T_D = \sum_{j=1}^{k} w_j \hat{T}^{(j)}$ and $1 = \sum_{j=1}^{k} w_j$, we get:

$\hat{T}_C - T_D = \frac{1}{k} \sum_{j=1}^{k} \hat{T}^{(j)} + \frac{\lambda_1}{k} \sum_{j=1}^{k} (\hat{T}^{(j)})^2 + \frac{\lambda_2}{k} \sum_{j=1}^{k} \hat{T}^{(j)}$ ; and $\frac{\lambda_1}{k} \sum_{j=1}^{k} \hat{T}^{(j)} + \lambda_2 = 0$. Solving and putting:

$$\lambda_1 = \frac{k(\hat{T}_C - T_D - \frac{1}{k} \sum_{j=1}^{k} \hat{T}^{(j)})}{\left\{ \sum_{j=1}^{k} (\hat{T}^{(j)})^2 - \frac{1}{k} (\sum_{j=1}^{k} \hat{T}^{(j)})^2 \right\}} = \frac{k(\hat{T}_p - \hat{T}_{kNN})}{\left\{ \sum_{j=1}^{k} (\hat{T}^{(j)})^2 - \frac{1}{k} (\sum_{j=1}^{k} \hat{T}^{(j)})^2 \right\}},$$

and $\lambda_2 = -\frac{\lambda_1}{k} \sum_{j=1}^{k} \hat{T}^{(j)}$ into (3), the required result (2) is obtained.

We note that the first condition of the Lemma ensures that $\hat{T}_C + T_B$ is calibrated to the data integrator, $\hat{T}_p$, of Kim and Tam (2021).

## 3.2    SMALL AREA ESTIMATORS WITH CkNN

It follows from the Lemma that the CkNN, or hybrid, estimator of the population total is given by:

$$\hat{T}_{CkNN} = \hat{T}_P$$

$$= T_B + T_D + \sum_{j=1}^{k} w_j \hat{T}^{(j)}$$

$$= T_B + T_D + \sum_{j=1}^{k} w_j \left( \sum_{i \in C \setminus D} y_{i(j)} \right)$$

$$= T_B + T_D + \sum_{i \in C \setminus D} \left( \sum_{j=1}^{k} w_j y_{i(j)} \right)$$

from which, the calibrated kNN small area estimator for small area $m$ is given by:

$$\hat{T}_{CkNN_m} = T_{B_m} + T_{D_m} + \sum_{i \in C_m \setminus D_m} \left( \sum_{j=1}^{k} w_j y_{i(j)} \right).$$ To simplify notations, we will use $\hat{T}_m$ to represent $\hat{T}_{CkNN_m}$ and

$\hat{T}_{C_m \setminus D_m}$ to represent $\displaystyle\sum_{i \in C_m \setminus D_m} \left( \sum_{j=1}^{k} w_j y_{i(j)} \right)$ in the sequel.

### 3.3 MEAN SQUARED ERROR FOR $\hat{T}_m$

From $\hat{T}_m = T_{B_m} + T_{D_m} + \hat{T}_{C_m \setminus D_m}$ and $T_m = T_{B_m} + T_{D_m} + T_{C_m \setminus D_m}$, we have

$$\hat{T}_m - T_m = \hat{T}_{C_m \setminus D_m} - T_{C_m \setminus D_m}$$

$$= \left\{ \hat{T}_{C_m \setminus D_m} - E(\hat{T}_{C_m \setminus D_m}) \right\} + \left\{ E(\hat{T}_{C_m \setminus D_m}) - T_{C_m \setminus D_m} \right\}$$

and, decomposing mean squared errors into variance and bias squared, we have

$$MSE(\hat{T}_m) = E(\hat{T}_m - T_m)^2$$

$$= E \left\{ \hat{T}_{C_m \setminus D_m} - E(\hat{T}_{C_m \setminus D_m}) \right\}^2 + \left\{ E(\hat{T}_{C_m \setminus D_m}) - T_{C_m \setminus D_m} \right\}^2$$

$$= E \left\{ \hat{T}_{C_m \setminus D_m} - E(\hat{T}_{C_m \setminus D_m}) \right\}^2 + E^2(\hat{T}_{C_m \setminus D_m}) e_m^2$$

where $e_m = \left\{ E(\hat{T}_{C_m \setminus D_m}) - T_{C_m \setminus D_m} \right\} / E(\hat{T}_{C_m \setminus D_m})$. The error, $E(\hat{T}_{C_m \setminus D_m}) - T_{C_m \setminus D_m}$, is due to the use of nearest neighbours to impute the missing values in $C_m \setminus D_m$ and is the imputation bias. We may describe $e_m$ as the relative imputation bias.

As the unobserved $E(\hat{T}_{C_m \setminus D_m})$, $Var(\hat{T}_m)$ and $e_m$ are functions of kNNs, it is intractable to give a closed form for them. For the first two quantities, the bootstrap offers an attractive technique to provide the estimates. However, Abadie & Imbens (2008) has shown that the "naïve" bootstrap (Efron and Gong, 1983) does not work for kNN problems. This is because the naïve bootstrap does not preserve the distribution of the number of times, denoted by $K_k(i)$, the unit $i \in D$, is used as a donor for imputing the missing values in $C \setminus D$. It is easily seen that $K_k(i)$ changes if bootstraps are drawn from $D$. To overcomes this problem, we use an approach inspired by Otsu and Rai (2017) who use a "fixed - $k$ asymptotic" bootstrap to estimate treatment effects in observational studies based on imputing the "counterfactuals" using kNN, which ensures that $K_k(i)$, $i \in D$, is not altered.

### 3.3.1 ESTIMATING $E(\hat{T}_{C_m \setminus D_m})$ and $Var(\hat{T}_m)$

It is easy to see that $\hat{T}_{C_m \setminus D_m} = \sum_{i \in A} y_i K_{km}(i)$, where $K_{km}(i) \geq 0$ is defined by $\sum_{j=1}^{k} n_{km}^{(j)}(i) w_j$ where $n_{km}^{(j)}(i)$ is

the number of times $y_i$ is used as the $j^{th}$ nearest neighbour for all the missing data points in

$C_m \setminus D_m$. Because by construction $w_j \approx \frac{1}{k}$, $K_{km}(i) \approx \frac{1}{k} \sum_{j=1}^{k} n_{km}^{(j)}(i)$. Let $z_{mi} = y_i K_{km}(i)$ $|A| = n$ and

$\psi_m = \{z_{m1}, ..., z_{mi}, ..., z_{mn}\}$. The following procedures select "fixed - $k$ asymptotic" bootstraps:

Step 1. Create a bootstrap, $\psi_m^*$, of the same size $n$, by sampling $\psi_m$ independently and with replacement.

Step 2. Repeat Step 1 $B$ times to create $\psi_{1m}^*, ..., \psi_{bm}^*, ..., \psi_{Bn}^*$, where $\psi_{bm}^* = \{z_{bm1}^*, ..., z_{bmi}^*, ..., z_{bmn}^*\}$.

Compute $\hat{T}_{C_m \setminus D_m b}^* = \sum_{i=1}^{n} z_{bmi}^*$ and $\hat{\bar{T}}_{C_m \setminus D_m} = \frac{1}{B} \sum_{b=1}^{B} \hat{T}_{C_m \setminus D_m b}^*$ .

Then $\hat{E}(\hat{T}_{C_m \setminus D_m}) = \hat{\bar{T}}_{C_m \setminus D_m}$ and $\hat{Var}(\hat{T}_m) = \hat{Var}(\hat{T}_{C_m \setminus D_m}) = \frac{1}{B} \sum_{b=1}^{B} (\hat{T}_{C_m \setminus D_m b}^* - \hat{\bar{T}}_{C_m \setminus D_m})^2$ . By treating $z_{mi}$ as

"observations" and resample them, the process is equivalent to resampling from $\{y_{mi}, \boldsymbol{x}_{mi}, K_{km}(i)\}_{i=1}^{n}$ (Otsu and Rai, 2017) and hence $K_{km}(i)$ is not altered by the bootstrap process.

How many fixed - $k$ asymptotic bootstraps are required? The bootstrap sample size B is important to determine the accuracy of the end points of the confidence interval. For accelerated bias correction confidence interval end-points, Efron (1987) shows that, $B \approx \left\{ \frac{1.71}{CV_W} \right\}^2$ , where $CV_W$ is the coefficient of variation of the "width" of the 95% bootstrap confidence interval. It is suggested we use this rule for determining the size of $B$ for variance estimation. For the numerical example below, we set $B = 500$, in order to get a $CV_W$ of about 7%.

### 3.3.2 ESTIMATING $e_m$

Let $\hat{e}_m$ be an estimator of the relative imputation bias, $e_m = \left\{ E(\hat{T}_{C_m \setminus D_m}) - T_{C_m \setminus D_m} \right\} / E(\hat{T}_{C_m \setminus D_m})$. Noting that because we know the true value of the target variable in $D$, we can proceed to estimate $e_m$ as follows:

Step 1.    Pick one data point, say $y_i$ from $D$. Use the HasD metric to find its $k$ nearest neighbours, $y_{i(1)}, y_{i(2)}, ..., y_{i(k)}$, from $D \setminus \{y_i\}$. Compute $\hat{y}_i = \sum_{k=i}^{K} w_k y_{i(k)}$. Create the pair $(y_i, \hat{y}_i)$.

Step 2.    Pick a second data point, $y_i$ from $D \setminus \{y_l\}$, where $l$ is different from previously selected data point(s). Use the HasD metric to find the $k$ nearest neighbours, $y_{l(1)}, y_{l(2)}, ..., y_{l(k)}$ from $D \setminus \{y_l\}$. Compute $\hat{y}_l = \sum_{k=i}^{K} w_k y_{l(k)}$. Create the pair $(y_l, \hat{y}_l)$.

Step 3.    Repeat Step 2 $n_{D_m}$ times to create the sets $\chi_m = \{(y_{mi}, \hat{y}_{mi}) : mi \in D_m\}, m = 1, ..., M$.

Step 4.    Compute an estimate of $e_m$ by $\hat{e}_m = \dfrac{\sum_{i \in \chi_m} (\hat{y}_{mi} - y_{mi})}{\sum_{i \in \chi_m} \hat{y}_{mi}}$

Thus, the bootstrap estimate of $E^2(\hat{T}_{C_m \setminus D_m}) e_m^2$ is given by $\hat{E}^2(\hat{T}_{C_m \setminus D_m}) \hat{e}_m^2$.

Step 4 (alternative).  When $\sum_{i \in D_m} \hat{y}_{mi}$ (for binary variables) or $n_{D_m}$ (for continuous variables) $\leq 5$, the estimate of $e_m$ becomes unstable.  In such situations, compute the alternative estimator of $\hat{e}_m$ by $\dfrac{\sum_{m=1}^{M} \sum_{i \in \chi_m} (\hat{y}_{mi} - y_{mi})}{\sum_{m=1}^{M} \sum_{i \in \chi_m} \hat{y}_{mi}}$.

## 4    AN AUSTRALIAN EXAMPLE

To illustrate our methods, we use the 1% public use micro data file from the 2016 Australian Census (Australian Bureau of Statistics, 2016) (available at https://www.abs.gov.au/statistics/microdata-tablebuilder/log-your-accounts to authorised users) to create the population, big data, and the probability sample.   Whilst it would be more preferable to use the 100% sample than the 1% sample from the Census, access to the 100% file by researchers is restricted due to privacy considerations.  Volunteers are defined in the 2016 Census as people who performed volunteer work for an organisation or group.  This consists of help willingly given in the form of time, service or skills, to a club, organisation or association in the twelve months prior to the 2016 Census.

The population, $U$, has 173,021 personal records.  With 56 regions, there is an average of 3,089 personal records per region.  Among the 173, 021 personal records, there was a total of 35,742 volunteers, giving an overall average volunteer participation rate of about 21%.  The number of volunteers ranged between 46 to 1,236 amongst the 56 small areas, and the volunteer participation rate varies between 11% to 31%.

For this example, we are interested to estimate the number of volunteers using both the big data and probability sample for 56 small areas as defined geographically by the Australian Bureau of Statistics.

A simple random sample of 1,730 (i.e. 1% of $U$ ) of the personal records was selected to form $A$. Amongst the 56 areas, the sample size ranges between 3 and 61, with the median at 28, and bottom and top 25% at 21 and 36 respectively. We grouped the small areas into regions, i.e. put contiguous areas together in the same regions such that we have roughly the same number of small areas in all three regions, having regard to Australia's State boundaries. Selection of the big data sample was conducted in the following manner:

    (i)        for Region 1, the sample comprised 50% of the volunteer personal records;
    (ii)      for Region 2, the sample comprised 50% of the non-volunteer personal records; and
    (iii)    for Region 3, the sample was taken randomly from 80% of the personal records.

In addition, we do not keep information on which of the big data records comes from which Region. Hence, mechanism for selecting the big data set is missing-not-at-random. The size of the big data sample using the above sampling scheme has 103,438 personal records (i.e. 60% of $U$ ) and 18,548 volunteers (52% of all volunteers). Leaving out the data points in $A \cap C$ which are observed, there are 68,908 data points in $C \setminus A$, to be imputed for volunteers. By comparison, the number of volunteers in $A$ is 341, which ranges between 0 and 14 in $A$ and between 0 and 8 in $A \cap C$ respectively amongst the 56 small areas. Using the method of Kim and Tam (2021), the asymptotically design-unbiased estimate of the total number of volunteers in the population is 36,312. This compares with the actual number of 35,742.

For the CkNN algorithm, we use the following auxiliary variables to find the nearest neighbours: labour force status (employed, unemployed and not in the labour force), birth region (6 groups), age (7 broad groups) and sex (male, female). We also used HasD as the distance metric. Furthermore, we give a value of 1 to a (predicted or actual) volunteer, and 0 otherwise. Thus

$L\left(y_{mi}, f^{-j(mi)}(\boldsymbol{x}_{mi}, k, p\right) = 1$ if $f^{-j(mi)}(\boldsymbol{x}_{mi}, k, p)$ gives a false positive, and $L\left(y_{mi}, f^{-j(mi)}(\boldsymbol{x}_{mi}, k, p\right) = -1$ for a false negative, prediction that the $mi^{th}$ unit is a volunteer. It follows from this definition of the loss function that the test error rate is the average net prediction errors amongst the 56 areas.

To determine the optimum $k$ and $p$, we used $K = 5$ to divide $D = A \cap C$ into 5 folds, and computed the objective function in (1) over all possible combinations (i.e. grid search) of $k = 1, \ldots 20$ and $p = 1, \ldots, 4$ to find the combination with the smallest test error rate. Note that for each $p$, there are $C_p^4$ combinations of the four auxiliary variables that may be used for feature selection. Hence the total number of all possible combinations to be tested with is 300.

Figure 1 is a plot of the test error rate against the different combinations of features, $p$ and $k$. These are presented in the horizontal axis of Figure 1, and description of each combination's features, $p$ and $k$ values are outlined in Appendix 1. It can be seen that the test error rate ranges from 19% to 33%. The lowest test error rate of 19% is achieved using the age, birthplace and labour force status variables, and $k = 5$. A test error rate of 19% suggests that the algorithm with this choice of $k$ and the features is expected to be "off" of the national total target by about the same amount. With calibration, however, this off target is substantially reduced, as shown in Table 1 to average at 8.9% across the 56 small areas.

Figure 1: Percentage of net prediction error by different features, $p$ and $k$

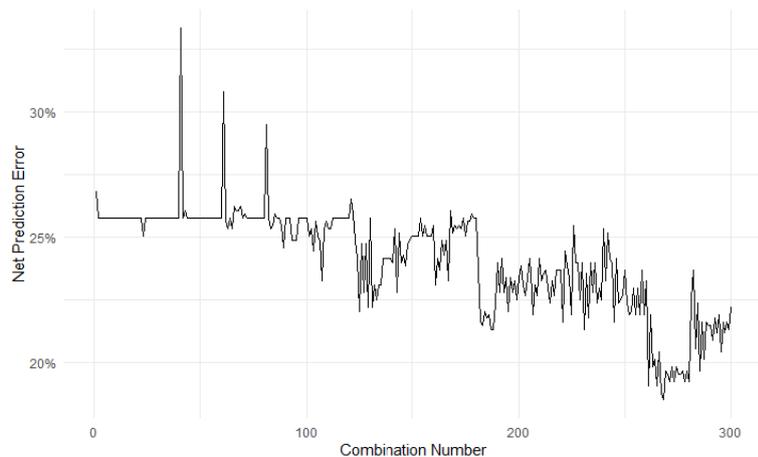

The hybrid estimates of volunteers for the 56 small areas, which we shall denote by $\hat{T}_m^{HY}$, and their mean squared errors $RT\hat{M}SE^{HY} = \sqrt{\hat{MSE}(\hat{T}_m^{HY})}$, are provided in the second column and third column respectively in Table 1 in Appendix 2.

The $\hat{T}_m^{HY}$ results in Table 1 are compared with $T_m$ in the upper left graph of Figure 2.

As an indication of the efficacy of the proposed method, we also applied the FH model to estimate the number of volunteers and their mean squared errors, using the same variables used to determine the nearest neighbours. Under the FH formulation, we have the sampling error model and the linking model defined respectively as follows:

Sampling model: $\hat{T}_m^{PR} = T_m + \varepsilon_m$, $\varepsilon_m$ is iid and has mean and variance of 0 and $\sigma_m^2$ respectively;

Linking model: $T_m = \boldsymbol{x}_m^T \boldsymbol{\beta} + u_m$, $u_m$ is iid and distributd as $N(0, \sigma_u^2)$

where $\varepsilon_m \perp u_m$ for $m = 1, ..., M$, and where $\hat{T}_m^{PR}$ represents the estimate of $T_m$ using a probability sample, the sampling variance $\sigma_m^2$ is assumed known, $\sigma_u^2$ and $\boldsymbol{\beta}$ are the (unknown) linking model variance and $p \times 1$ vector of regression coefficients respectively (Fay and Harriot, 1979; Molina and Marhuenda, 2015). Also $\boldsymbol{x}_m$ is the vector of totals of the same auxiliary variables as those used for computing the HasD metric for the nearest neighbours above, i.e. labour force status (employed, unemployed and not in the labour force), birth region (6 groups) and age (7 broad groups, for the $m^{th}$ small area. The empirical best linear unbiased predictor for SAE for the $m^{th}$ area is

$\hat{T}_m^{FH} = \hat{\gamma}_m \hat{T}_m^{PR} + (1 - \hat{\gamma}_m) \boldsymbol{x}_m^T \hat{\boldsymbol{\beta}},$

where $\hat{\gamma}_m = \dfrac{\hat{\sigma}_u^2}{\hat{\sigma}_u^2 + \sigma_m^2}$ and $\hat{\boldsymbol{\beta}} = \left\{ \displaystyle\sum_{m=1}^M (\hat{\sigma}_u^2 + \sigma_m^2)^{-1} \boldsymbol{x}_m \boldsymbol{x}_m^T \right\} \left\{ \displaystyle\sum_{m=1}^M (\hat{\sigma}_u^2 + \sigma_m^2)^{-1} \boldsymbol{x}_m \hat{T}_m^{PR} \right\}.$

The FH estimates together with their MSEs are summarised in the last two columns of Table 1. The calculations were carried out using the **SAE** package (Molina and Marhuenda, 2015). Note that normality assumption is not required for point estimation.

The $\hat{T}_m^{FH}$ results in Table 1 are compared with $T_m$ in the upper right graph of Figure 2.

It can be concluded from Figure 2 that the hybrid estimates, $\hat{T}_m^{HY}$, are more accurate (closer to the diagonal line) and precise (narrow error bands) than the FH estimates, $\hat{T}_m^{FH}$. Specifically, we can see from Table 1 that the average absolute estimation error (AAER), average relation root mean squared error (ARRTMSE), and coverage rate for the hybrid estimates are 57, 11% and 93% respectively as compared with 107, 28% and 93% respectively for the FH estimates.

In comparing the precision of these estimates, it should be noted that the inference framework for hybrid estimates is designed-based, and for FH estimates model-based.

As the FH model is an area model and given that unit record data is available, we are doing further work to compare estimates from our method with the EBLUP using the unit level model of Battese et. al (1988). We plan to use the data in $B \cup (A \cap C)$ as observed unit data as covariates and a unit-level mixed models (Hobza and Morales, 2016). We hope to publish the results of this research in a future paper.

As note earlier, a number of researchers have suggested to use big data directly as auxiliary variables for FH estimation. As far as we are aware, no work has been carried out to date to assess how the differential under-coverage rates in the small areas may affect the FH estimates. To assess this, we have conducted two experiments by artificially creating differential under-coverage rates of the population in the 56 areas by deleting certain number of personal records before re-running the **SAE** package. The deletion rates used for these areas are as follows:

Experiment 1 - 5% of the Age1 group records in areas 1- 8 deleted; 10% of the Age2 group records in areas 9-16 deleted; 15% of Age3 group records of the 17-24 area deleted; 20% of Age4 group records in areas 25-32 deleted; 25% of Age5 group records in areas 33 – 40 deleted; 30% of Age6 group records in areas 41-48 deleted and for the rest, 35% of Age 7 group records deleted. The results are denoted by $\hat{T}_m^{FH^{(1)}}$ and $RTM\hat{S}E^{FH^{(1)}} = \sqrt{M\hat{S}E(\hat{T}_m^{FH^{(1)}})}$ in Table 2 in the Appendix.

Experiment 2 - 35%, 40%, 45%, 50%, 55%, 60% and 65% of the records in areas 1- 8, 9-16, 17-24, 25-32 , 33 – 40, 41-48 and 49-56 deleted respectively. The results are denoted by $\hat{T}_m^{FH^{(2)}}$ and $RTM\hat{S}E^{FH^{(2)}} = \sqrt{M\hat{S}E(\hat{T}_m^{FH^{(2)}})}$ in Table 2 in the Appendix.

The data from both experiments are compared against $T_m$ in the lower second two graphs of Figure 2.

Figure 2: Plot of $\hat{T}_m, \hat{T}_m^{FH}, \hat{T}_m^{FH^{(1)}}, \hat{T}_m^{FH^{(2)}}$ and their error bars against $T_m$

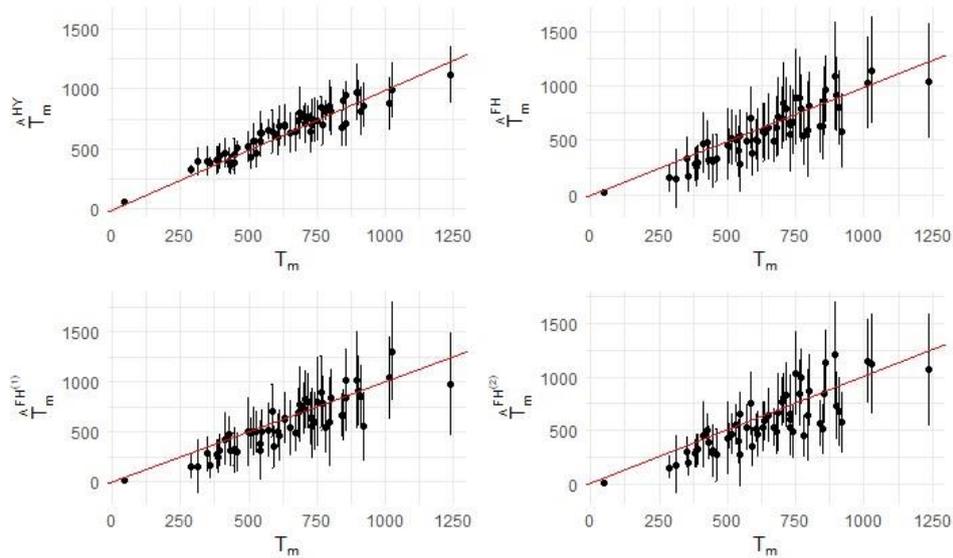

Data from Table 1 (upper left and upper right); and data from Table 2 (lower left and left right). Error band is defined as

$$\hat{T}_m^I \pm 1.96\sqrt{M\hat{S}E(\hat{T}_m^I)}\ )\ \text{where}\ \ I = HY, FH, FH^{(1)}\ \text{or}\ FH^{(2)}.$$

The lower two graphs of Figure 2 show that the accuracy (as measured by AAER) of the FH estimates is affected to a different extent by the different coverage rate of the auxiliary data. In this numerical example, whilst the reduction in accuracy and precision is marginally affected by incomplete records in the small areas (Experiment 1), it is more significantly affected by incomplete auxiliary information (Experiment 2). This result is in accordance with intuition as there is more information available in Experiment 1 than Experiment 2 for the FH modelling. On the other hand, the estimated coverage rates in both experiments are not statistically significant from the nominal coverage rate of 95% (at 95% confidence).

## 5. DISCUSSION AND CONCLUSION

This paper outlines a hybrid estimation methodology using calibrated k nearest neighbours for small area estimation. The idea is to borrows strength from big data for SAE. Pre-requisites for the methodology to work are:

a) the target variables are observed throughout the big data set without measurement errors – this condition is more likely to be satisfied when using administrative data for hybrid estimation than using many other types of big data. Where this condition is not satisfied, $A \cap B$ can be used as a training data set to construct a measurement error model to adjust the target variables in big data (Medous et. al., 2022);

b) there are no over-coverage errors in the big data. Where this is not the case, $A_m \cap B_m$ can be used to estimate over-coverage rates in the small areas to remove the bias from $T_{B_m}$ ;

c) the donor set, $D$, which depends on the size of $B$ and $A$ has to be sufficiently large, to support the imputations. As to what exactly should the size be to make hybrid estimation worthwhile is a topic for further research. In the numerical example, there are 174 donor volunteers out of 675 personal records to support about 69,000 imputations. It is also noted in the numerical example that the imputation bias is the dominant contributing factor of the MSE. This suggests that choice of auxiliary variables and a suitable distance metric for determining the nearest neighbours are also important considerations to minimise the imputation bias;

d) $\delta_m$ is fully observed for the units in $A$ - this can generally be made possible by matching the units between $A$ and $B$ through direct matching or probability matching (Fellegi and Sunter, 1969). When $\delta_m$ is not observed, or observed with error, Kim and Tam (2021) developed, using a semi-supervised classification technique, an EM estimator for $\delta_m$ ;

e) associated with each unit of the population, there is a set of covariates which are available and known to the statistician. The assumption presumes the existence of a database with covariates covering the whole of the population. Do such databases exist? They exist (in the form of population registers) in the Scandinavian countries. In New Zealand, a large research database called Integrated Data Infrastructure (IDI) (Statistics New Zealand, 2022) is constructed and maintained using records from government agencies, Statistics New Zealand surveys and non-government organisations. The IDI holds de-identified microdata about people and households in New Zealand and contains such life events information as education, income, benefits, migration, justice and health. In Australia, the multi-agency data integration project (Australian Bureau of Statistics, 2015) created a secure database that integrates records from Australian Bureau of Statistics, Australian Taxation Office, Department of Education, Department of Health and Aged Care, Department of Social Services, Service Australia and Department of Home Affairs. In addition, it is customary for national statistical offices to maintain sampling frames from which probability samples are drawn for business or household surveys. Such sampling frames normally has a limited set of covariates to assist with the selection of optimal samples. For example, business sampling frames generally contain information on the geographic location, industry, employment size in broad categories of businesses; and

f)   the big data set has a target variable of interest to the official statistician, and the national statistics office has a probability survey that collects the same variable, e.g. employment and unemployment data collected from online panels (Callegaro and DiSogra, 2008), and labour force surveys conducted by the national statistics office. The case when the target variable in the big data set suffers from measurement errors has been dealt with in a) above.  Integrating data from online panels and probability surveys is the most promising way to construct hybrid estimates.

Note that whilst a calibrated ensemble of machine learning methods may be used in lieu of the CkNN method of imputation, i.e. $\hat{T}_{EN_m} = T_{B_m} + T_{D_m} + \sum_{j=1}^{k} w_j \hat{T}_{ML_m}^{(j)}$ where $\hat{T}_{EN} = \sum_{m=1}^{M} \hat{T}_{EN_m} = \hat{T}_P$ and $\hat{T}_{ML}^{(j)}$ denotes the $j^{th}$ machine learning (ML) method which may be identical with one another but with a different hyper-parameter (e.g. nearest neighbours algorithm, but each ML method differs by their ranking in terms of nearest neighbours) or different in the method itself (e.g. using 1NN for $j = 1$ and support vector machines for $j = 2$ etc.), we prefer to use jNN for the ML algorithm for $j = 1, ..., k$ because the nearest neighbour methodology has the advantage of minimising the Expected Prediction Error under quadratic loss (Hastie et al, 2008, p. 18).

Using Australian population census data, the CkNN, or hybrid, estimator developed in this paper was compared with FH estimates using an off-the-shelf package and we found that our estimator is on average more accurate and precise than the FH estimator.  As noted by a referee, however, the FH estimates in the numerical example are based on an off-the-shelve model but can be improved by, for example, including interactions between the area level covariates in the linking model, or using a binomial likelihood combined with a beta or logit-normal model for the probability of volunteering.

We have also conducted experiments on the accuracy and precision of the FH estimates using auxiliary variables that come from a big data set subject to under-coverage error.  The numerical example shows that accuracy of the FH estimates is more affected by under-coverage of personal records in the small areas than by under-coverage of auxiliary variables in the personal records.  Ybarra and Lohr (2008) provided methods to adjust for FH estimates when the auxiliary information is subject to sampling variation.

Provided that the pre-requisites underpinning hybrid estimation are satisfied,  hybrid estimation, which borrow strength from a suitable big data source, (a) is relatively assumptions free – even though they rely on CkNN to impute the missing values, the variance due to imputation bias can be estimated using a design-based method – and requires less effort to implement e.g. no effort required to develop and test the linking model otherwise required for FH estimation; (b) enables consistency between the sum of SAEs and the national estimate of the variable of interest, thus not undermining confidence in official statistics; (c) through calibration, can render the SAEs to be more accurate than the FH estimates; (d) allows a design-based estimate of the MSE; and (e) is target variable agnostic, i.e. the same algorithm can be applied to target variables of different character. Finally, whilst the CkNN method outlined in this paper looks promising as a design-based estimation method, more work needs to be done to assess its efficacy against EBLUP derived from unit level modelling.

**Acknowledgement**

We would like to thank an Associate Editor and two referees for their helpful comments on an earlier version of the paper. An earlier version of this paper was presented to the Methodology Advisory Committee of the ABS and we would like to thank the Committee members for their comments as well. The views expressed in this paper are those of the authors and do not necessarily represent those of the ABS.

**APPENDIX 1**

The 300 combinations in Figure 1 comprises blocks of 20 combinations each of which consists of k = 1,…,20 and one feature below in the following order.  E.g. the first combination comprises SEX and k=1 and the 20th combination comprises SEX and k=20.

**Feature details**

SEX

Labour Force Status (LFS)

LFS x SEX

Birth Region (BR)

BR x SEX

BR x LFS

BR x LFS x SEX

AGE

AGE x SEX

AGE x LFS

AGE x LFS x SEX

AGE x BR

AGE x BR x SEX

AGE x BR x LFS

AGE x SEX x LFS x BR

**APPENDIX 2**

**Table 1: Estimates for** $\hat{T}_m$ , $\hat{T}_m^{HY}$ , $RTM\hat{S}E^{HY}$ , $\hat{T}_m^{FH}$ **and** $RTM\hat{S}E^{FH}$

| Small area | $T_m$ | $\hat{T}_m^{HY}$ | $RTM\hat{S}E^{HY}$ | $\hat{T}_m^{FH}$ | $RTM\hat{S}E^{FH}$ |
|---|---|---|---|---|---|
| 1 | 910 | 811 | 104 | 800 | 176 |
| 2 | 426 | 401 | 55 | 478 | 105 |
| 3 | 728 | 641 | 83 | 642 | 141 |
| 4 | 1014 | 879 | 111 | 1029 | 216 |
| 5 | 839 | 676* | 78 | 635 | 127 |
| 6 | 383 | 404 | 59 | 276 | 80 |
| 7 | 529 | 468 | 62 | 511 | 112 |
| 8 | 730 | 741 | 88 | 668 | 174 |
| 9 | 447 | 387 | 55 | 308 | 129 |
| 10 | 433 | 364 | 43 | 317 | 95 |
| 11 | 544 | 630 | 93 | 280* | 134 |
| 12 | 751 | 733 | 87 | 898 | 225 |
| 13 | 1236 | 1119 | 122 | 1046 | 272 |
| 14 | 650 | 636 | 87 | 623 | 156 |
| 15 | 350 | 400 | 62 | 332 | 100 |
| 16 | 499 | 521 | 64 | 458 | 169 |
| 17 | 312 | 392 | 58 | 146 | 142 |
| 18 | 768 | 701 | 88 | 790 | 158 |
| 19 | 507 | 436 | 52 | 439 | 83 |
| 20 | 857 | 714 | 91 | 972 | 156 |
| 21 | 1026 | 990 | 119 | 1144 | 252 |
| 22 | 732 | 698 | 73 | 559 | 173 |
| 23 | 706 | 708 | 86 | 848 | 187 |
| 24 | 584 | 646 | 91 | 704 | 148 |
| 25 | 412 | 462 | 62 | 474 | 142 |
| 26 | 800 | 813 | 101 | 821 | 156 |
| 27 | 896 | 968 | 121 | 1098 | 253 |
| 28 | 794 | 859 | 110 | 594 | 226 |
| 29 | 391 | 438 | 56 | 294 | 82 |
| 30 | 607 | 596 | 76 | 502 | 95 |
| 31 | 632 | 698 | 90 | 583 | 128 |
| 32 | 543 | 561 | 65 | 404 | 117 |
| 33 | 920 | 861 | 95 | 586 | 176 |
| 34 | 445 | 457 | 65 | 317 | 92 |
| 35 | 670 | 641 | 84 | 490 | 103 |
| 36 | 685 | 798 | 110 | 719 | 187 |
| 37 | 741 | 739 | 94 | 665 | 153 |
| 38 | 387 | 406 | 56 | 257 | 95 |
| 39 | 515 | 563 | 78 | 522 | 131 |
| 40 | 611 | 692 | 87 | 497 | 149 |
| 41 | 589 | 639 | 37 | 386 | 112 |
| 42 | 459 | 504 | 32 | 335 | 111 |
| 43 | 898 | 970 | 54 | 920 | 175 |
| 44 | 570 | 657* | 43 | 494 | 133 |
| 45 | 633 | 690 | 41 | 571 | 137 |
| 46 | 847 | 904 | 46 | 634 | 159 |
| 47 | 702 | 770 | 46 | 709 | 142 |
| 48 | 681 | 740 | 44 | 620 | 143 |
| 49 | 763 | 844 | 48 | 893 | 188 |
| 50 | 716 | 774 | 44 | 789 | 150 |
| 51 | 548 | 637* | 40 | 545 | 120 |
| 52 | 359 | 390 | 23 | 164* | 67 |
| 53 | 853 | 948 | 59 | 864 | 164 |
| 54 | 289 | 324 | 22 | 152* | 67 |
| 55 | 779 | 818 | 43 | 539 | 135 |
| 56 | 46 | 55* | 4 | 20* | 13 |
| Total | 35,742 | 36,312 | - | 32,361 | - |
| Average absolute estimation error (AAEE) | - | 57 | - | 107 | - |
| Average relative root mean squared error (ARRTMSE) | - | - | 11% | - | 28% |
| Estimated coverage rate | - | 93% | - | 93% | - |

Notes: (1) $RTM\hat{S}E^I = \sqrt{M\hat{S}E(\hat{T}_m^I)}, I = HY, FH$ and the alternative estimator of $\hat{e}_m$ in Step 4 is used for $M\hat{S}E(\hat{T}_m^{HY})$

(2) * denotes $T_m$ is not within $\hat{T}_m^I \pm 1.96\sqrt{MSE(\hat{T}_m^I)}, I = HY, FH$

(3) $RRTM\hat{S}E^I = \sqrt{M\hat{S}E(\hat{T}_m^I)} / \hat{T}_m^I, I = HY, FH$

(4) Estimated coverage rate = (# of true counts within the 95% confidence interval) divided by 56. The coverage rate of 93% is not significantly different (95% confidence) from the nominal coverage rate of 95%.

**Table 2: FH estimates based on different big data coverage**

| Small area | $T_m$ | $\hat{T}_m^{FH}$ | $RTM\hat{S}E^{FH}$ | $\hat{T}_m^{FH^{(1)}}$ | $RTM\hat{S}E^{FH^{(1)}}$ | $\hat{T}_m^{FH^{(2)}}$ | $RTM\hat{S}E^{FH^{(2)}}$ |
|---|---|---|---|---|---|---|---|
| 1 | 910 | 800 | 176 | 854 | 156 | 673 | 167 |
| 2 | 426 | 478 | 105 | 478 | 90 | 504 | 79 |
| 3 | 728 | 642 | 141 | 639* | 133 | 529 | 132 |
| 4 | 1014 | 1029 | 216 | 1042 | 210 | 1147 | 202 |
| 5 | 839 | 635 | 127 | 666 | 109 | 570* | 110 |
| 6 | 383 | 276 | 80 | 277 | 76 | 284 | 69 |
| 7 | 529 | 511 | 112 | 500 | 105 | 555 | 98 |
| 8 | 730 | 668 | 174 | 563 | 167 | 650 | 191 |
| 9 | 447 | 308 | 129 | 313 | 118 | 310 | 125 |
| 10 | 433 | 317 | 95 | 318 | 85 | 384 | 80 |
| 11 | 544 | 280* | 134 | 309* | 140 | 270 | 152 |
| 12 | 751 | 898 | 225 | 800 | 227 | 1032 | 199 |
| 13 | 1236 | 1046 | 272 | 976 | 262 | 1066 | 268 |
| 14 | 650 | 623 | 156 | 538 | 142 | 644 | 128 |
| 15 | 350 | 332 | 100 | 281 | 89 | 295 | 76 |
| 16 | 499 | 458 | 169 | 504 | 172 | 428 | 176 |
| 17 | 312 | 146 | 142 | 147 | 138 | 179 | 141 |
| 18 | 768 | 790 | 158 | 782 | 138 | 996 | 132 |
| 19 | 507 | 439 | 83 | 483 | 78 | 443 | 62 |
| 20 | 857 | 972 | 156 | 1019 | 156 | 1131 | 155 |
| 21 | 1026 | 1144 | 252 | 1304 | 250 | 1124 | 237 |
| 22 | 732 | 559 | 173 | 633 | 153 | 607 | 168 |
| 23 | 706 | 848 | 187 | 823 | 148 | 799 | 117 |
| 24 | 584 | 704 | 148 | 712 | 136 | 751 | 152 |
| 25 | 412 | 474 | 142 | 425 | 134 | 454 | 157 |
| 26 | 800 | 821 | 156 | 840 | 143 | 871 | 170 |
| 27 | 896 | 1098 | 253 | 1014 | 247 | 1209 | 249 |
| 28 | 794 | 594 | 226 | 598 | 226 | 639 | 218 |
| 29 | 391 | 294 | 82 | 312 | 82 | 320 | 73 |
| 30 | 607 | 502 | 95 | 508 | 87 | 519 | 66 |
| 31 | 632 | 583 | 128 | 632 | 131 | 526 | 116 |
| 32 | 543 | 404 | 117 | 386 | 112 | 400 | 103 |
| 33 | 920 | 586 | 176 | 551* | 174 | 572* | 143 |
| 34 | 445 | 317 | 92 | 325 | 82 | 290 | 99 |
| 35 | 670 | 490 | 103 | 491 | 97 | 527 | 90 |
| 36 | 685 | 719 | 187 | 773 | 190 | 663 | 188 |
| 37 | 741 | 665 | 153 | 602 | 147 | 485* | 124 |
| 38 | 387 | 257 | 95 | 251 | 87 | 307 | 83 |
| 39 | 515 | 522 | 131 | 504 | 122 | 465 | 99 |
| 40 | 611 | 497 | 149 | 461 | 129 | 464 | 109 |

| 41 | 589 | 386 | 112 | 350* | 106 | 347* | 95 |
|---|---|---|---|---|---|---|---|
| 42 | 459 | 335 | 111 | 298 | 101 | 268 | 128 |
| 43 | 898 | 920 | 175 | 909 | 182 | 730 | 160 |
| 44 | 570 | 494 | 133 | 519 | 137 | 528 | 111 |
| 45 | 633 | 571 | 137 | 620 | 120 | 589 | 93 |
| 46 | 847 | 634 | 159 | 662 | 130 | 512* | 120 |
| 47 | 702 | 709 | 142 | 729 | 131 | 771 | 111 |
| 48 | 681 | 620 | 143 | 688 | 127 | 485 | 107 |
| 49 | 763 | 893 | 188 | 901 | 181 | 847 | 156 |
| 50 | 716 | 789 | 150 | 804 | 140 | 827 | 156 |
| 51 | 548 | 545 | 120 | 497 | 116 | 657 | 102 |
| 52 | 359 | 164* | 67 | 160* | 65 | 202 | 60 |
| 53 | 853 | 864 | 164 | 838 | 169 | 847 | 147 |
| 54 | 289 | 152* | 67 | 157* | 62 | 153* | 55 |
| 55 | 779 | 539 | 135 | 548 | 131 | 446* | 100 |
| 56 | 46 | 20* | 13 | 23 | 12 | 10* | 11 |
| Total | 35,742 | 32,361 | - | 32,337 | | 32,301 | - |
| Average absolute estimation error (AAER) | - | 107 | - | 109 | | 139 | - |
| Average relative root mean squared error (ARRTMSE) | - | - | 28% | - | 27% | - | 26% |
| Estimated coverage rate | - | 93% | - | 89% | - | 86% | - |

Notes: (1) * denotes $T_m$ is not within $\hat{T}_m^{FH} \pm 1.96\sqrt{M\hat{S}E(\hat{T}_m^{FH})}$ or $\hat{T}_m^{FH^{(i)}} \pm 1.96\sqrt{M\hat{S}E(\hat{T}_m^{FH^{(i)}})}, i = 1, 2.$

(2) The coverage rates of 93%, 89% and 86% are not significantly different (95% confidence) from the nominal coverage rate of 95%.